# Prediction of Reduced Glass Transition Temperature using Machine Learning

\* Prakash P [1], Akash Ravi [1], and Kailashnath N [2]

Department of Computer Science and Engineering,
Amrita School of Engineering, Coimbatore,
Amrita Vishwa Vidyapeetham, India

```
1* p_prakash1@cb.amrita.edu
1 cb.en.h4cse16001@cb.students.amrita.edu
2 cb.en.h4cse16003@cb.students.amrita.edu
```

\* Corresponding Author

**Abstract.** The advent of computational material sciences has paved the way for data-driven approaches for modeling and fabrication of materials. The prediction of properties like the glass-forming ability (GFA) by using the variation in alloy composition remains to be a challenging problem in the field of material sciences. It also results in significant financial concerns for the manufacturing industry. Despite the existence of various empirical guides for the prediction of GFA, a comprehensive prediction model is still highly desirable. This work focuses on studying some of the popular machine learning algorithms for the prediction of the reduced glass transition temperature ($T_{rg}$) of material compositions. From the experimentation, we conclude that the ensemble model performs better for predicting $T_{rg}$. This result can prove instrumental in the branch of material sciences by helping us to develop materials having remarkable properties.

## 1   Introduction

Estimation of alloys properties with respect to their glass-forming ability has been a field of interest in the exploration of bulk metallic glasses. Glass transition refers to the process of a gradual and reversible transition in amorphous materials from a hard and brittle state into a relatively viscous state. The transition temperature of a material characterizes the temperature range over which this transition occurs. The reduced glass transition temperature ($T_{rg}$) is formally defined to be the ratio of the transition temperature to the melting temperature of an alloy. In general, the increase in the glass-forming ability (GFA) is directly proportional to the increase in reduced glass transition temperature. There have been various studies [1 - 4] to experimentally demonstrate that there exists a correlation between the GFA and $T_{rg}$ values.

The quantitative study of these properties aids in the development of simple methods for making composites of nanoparticles and bulk metallic glasses. Being able to predict their formation from known parameters will help us easily leverage their distinctive combination of mechanical properties and plastic-like processability for potential applications [5]. A variety of commercial applications that would benefit from this research includes products catering to the needs of sectors like electronics, medicine, sports, and defense. These interdisciplinary applications and engineering activities ultimately depend upon materials to develop tangible products.

Machine learning techniques can help us create regression models without explicitly programming them. Training models on datasets allows them to learn a specific task through experience [6]. ML algorithms have been used to develop predictive models that support effective decision making for quite a while. In the domain of material sciences, these computer-aided techniques can be used to identify patterns in a data set [7, 8]. Consequently, it is possible to predict the value of a target variable, given the other attribute values. The effectiveness of such techniques can be evaluated based on metrics like mean absolute percentage error (MAPE).

This work researches various machine learning techniques applied to a dataset containing metallic glass descriptors. The performance of these models is comprehensively investigated and compared. The organization of the remaining sections is as follows: Section 2 surveys on existing research whose inferences are aligned with our work. Section 3 explains the dataset features and pre-processing techniques employed. Section 4 describes the setup used for the comparative study and evaluation for different methods. The results obtained from the ML models have been furnished in Section 5. Finally, the conclusions are presented in Section 6.

## 2      Related Work

There has been quite a lot of attempts to use machine learning techniques in the field of material sciences and help discover latent relationships and simulate processes. The results can potentially pave a path for innovation in existing industrial processes and may lead to promising directions for research.

Jonathan Schmidt et al. [9] explained the increasing potential of computers in identifying the patterns that govern the behavior of materials and their physical properties. The authors advocated the fact that computational methods can help researchers discover novel materials without conducting physical experiments that are often expensive. However, there has been a strong emphasis on the amounts of structured high-quality data that is critical to developing an effective model. Ramachandran K I et al. [10] provided a comprehensive overview of the field of computational chemistry and molecular modeling. The textbook illustrates the use of various visualization techniques and software tools for application in related fields like drug synthesis, polymer sciences, and biomedical engineering.

Similarly, Keith T. Butler et al. [11] summarized the recent progress in artificial intelligence and machine learning techniques that are increasingly relevant with respect to material sciences. Rather than relying upon intuitions, machine learning lets us mathematically deduce its predictions. In certain cases, these methodologies can be backtracked to identify important observations. Conclusively, there is a great untapped potential for new discoveries in this discipline.

With respect to the glass-forming ability of materials, Z.P. Lu et al. [3, 6] established the correlation between the $T_{rg}$ and the GFA based on experimental results. Xiao-Jin Xu et al. [4] also presented a detailed analysis of the specific roles of $T_{rg}$ and the nuances that might arise in characterizing the GFA. The scope of these researches does not include the usage of ML to do any prediction of indicators or the classification of materials based on their domain knowledge.

Y. T. Sun et al. [12] demonstrated a machine learning approach for the understanding and prediction of GFA. The author had used an SVM to predict the GFA of alloys using different input features. In a similar attempt, Daniel R. Cassar et al. [13] had documented the usage of artificial neural networks to predict the glass transition temperatures. Here, the discussions were primarily about the result of a single algorithm alone and this highlights the need for a comprehensive evaluation of multiple techniques and algorithms that can be used for this purpose.

Based on these works, it can be concluded that it is desirable to be able to predict the GFA of materials. Furthermore, this can effectively be translated to predicting the $T_{rg}$ of materials. By creating a model to estimate $T_{rg}$ for an arbitrary alloy, it could be possible to use the prediction to directly estimate GFA. Alternatively, it can also be fed as an input for another model to predict the GFA. We thus harness the potential of machine learning for creating models for the same and compare each of them.

## 3 Exploratory Data Analysis

### 3.1 Dataset

The dataset used for this research contains details on the material composition features and the reduced glass transition temperature. The data was assembled primarily by Vanessa Nilsen under the guidance of Prof. Dane Morgan at the University of Wisconsin-Madison [14]. The dataset uses a diverse list of attributes that are suitable for describing a wide variety of physical and chemical properties. The data has been partitioned into groups of closely related materials to increase predictive accuracy [15]. Data preparation is essential to apply various supervised learning models due to the existence of a target variable in the dataset.

The different attributes upon which the prediction is made includes the density composition average, details of periodic properties, valence composition average, the heat of vaporization, the difference between the boiling temperatures, specific heat capacity, etc. In addition to these columns, information about the majority element for each alloy has been included. This could turn out to be an interesting feature while trying to evaluate the importance of each property.

### 3.2    Pre-processing

Initially, different statistics are obtained to describe metrics like dispersion, central tendency, and shape of the dataset's distribution. For numerical data, the result's index would include the count, average, standard deviation, along with the lower, median and upper percentiles. Distribution of the data can be explored by plotting histograms for each attribute. Given the values of all series in a data-frame, this method groups them into bins and draws all of these bins on a single axis.

There are a lot of problems that could arise when handling data in the higher dimensions that are not as pronounced in lower dimensions. To reduce the dimension of the input data and select only the minimum number attributes, while still retaining most of the information, the correlation between attributes are calculated. Based on this result, all the highly correlated values are listed. Figure 1 depicts the correlation matrix generated for the dataset.

Variance Threshold is a baseline approach used to select an optimal set of features. All the features where the variance along a column does not exceed the threshold value are dropped from further processing. Care is taken to ensure that the majority of data can still be explained by the residual attributes. Based on these methods, the number of features is effectively reduced to 13 from 20. In the context of material properties, the features that have been pulled out of the data often signify the dependencies and the potential inferences that can be made as a result of them. As an example, a few properties averaged over the material composition and the features that are only for the majority element in each alloy can often be inferred from each other.

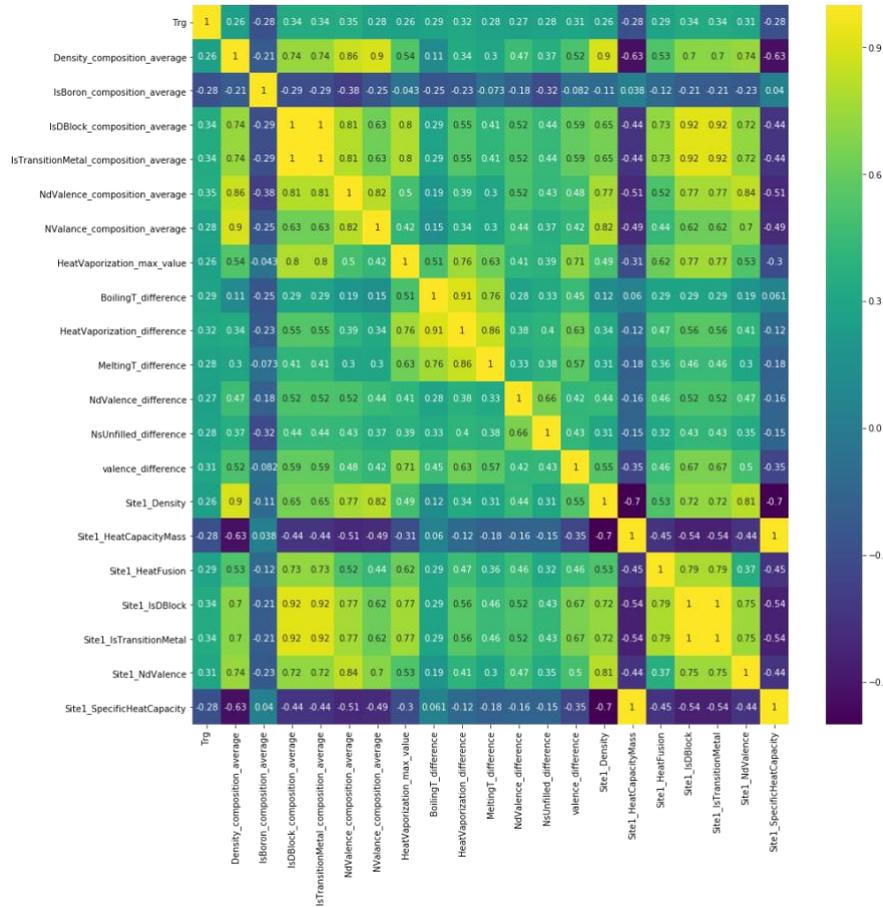

**Fig. 1.** Correlation heatmap between features

Post dimensionality reduction, the possibility to fit standard distribution functions to the attributes is evaluated. Kernel density estimation (KDE) is a technique to analyze and estimate a random variable's probability density function (PDF). It creates a smooth curve given a dataset. Upon using this procedure, it can be found that a few attributes can be fit into a combination of normal distributions, while a few attributes do not seem to follow any well-defined distribution function.

As an attempt to remove the outliers, box plots are made for each attribute to graphically depict groups of numerical data using their quartiles. Univariate outliers are analyzed by looking for points that are past the end of the whiskers in the box plots. For multivariate outlier analysis, the data is re-scaled and centered by calculating Z-scores. The data points which are way too far from the origin are treated as outliers and need not be considered during the processing. In our case, a threshold of 3 or -3 is used to filter them. Most of the data points are observed to fall between this range.

The histograms are checked once again to ensure that the data is consistent after all these pre-processing. Good knowledge of the data can help us develop an intuition for interpreting the results of future processing and analysis. We thus generate hypotheses for further analysis and prepare the data for training our ML algorithms.

## 4       Proposed Architecture

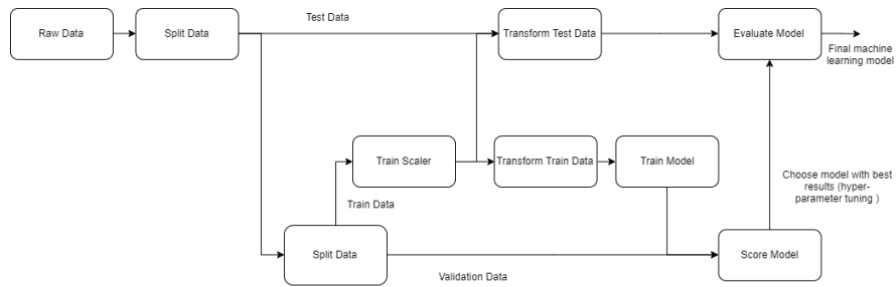

**Fig. 2.** Proposed Machine Learning Pipeline

Headings After all the pre-processing has been done, the data is split into random train and test subsets. 15% of the data has been allocated for testing, while the rest will be utilized for training the model. Figure 2 explains the architecture and the flow of the data processing pipeline used in this paper. K-Fold cross-validation has been used for assessing the ability of models to generalize to fit an independent data set. This method provides indices to split data into different sets for testing and training. It splits the dataset into k consecutive folds. Each fold is used once for validation, while the remaining k - 1 folds form the training set. The K value is set to be 12 to balance the bias-variance trade-off [16].

A hyperparameter is a characteristic value of a model that is external to it and can't be estimated from the data. The hyperparameter value must be set before the start of the learning process. Grid search is a technique to find optimal values of hyperparameters by building a model for every combination specified and evaluating them.

To check if creating machine learning models for this prediction task is really useful, the mean absolute percentage error (MAPE) of each model is compared against a naive prediction, i.e. constantly predicting the mean of the test data as a result, irrespective of the input data points. Any model performing worse than this is not effective. The prediction accuracy of our forecasting methods is thus measured.

When approaching any machine learning problem, it is possible to choose amongst many different algorithms and techniques. Regardless of guidelines that can help us select a model, usually, the best way to determine the most suitable one is to test our algorithms directly through trial and error. For the purpose of experimenting and performing an interactive analysis of the data, libraries like Scikit learn, NumPy, Matplotlib and Seaborn were used in Jupyter Notebooks. The code was executed in a cloud computing environment for better performance and collaborative research. The dataset is thus used for training against a wide range of ML techniques including multi-variable linear regression, Bayesian ridge regression, SVM, Regression trees, and Ensemble models.

## 5   Results

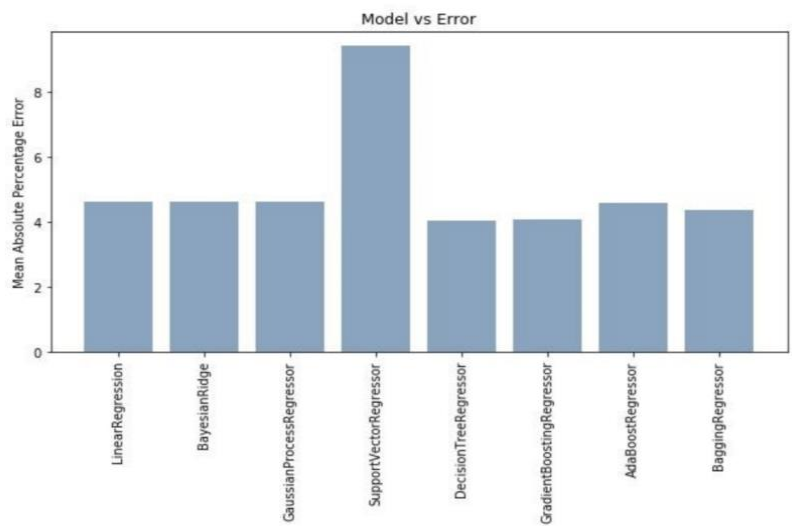

**Fig. 3.** Bar plot describing error rates for various model

For each ML algorithm chosen, the model is trained and the MAPE is calculated. Based on the results collected from analyzing the performance of all these ML models, it can be conclusively noticed that both the regression tree model and the ensemble model perform better when compared to other models. Figure 3 outlines the error rates for each of these models and the MAPE values are listed in table 1. Regardless of the slight variations in the results that would be predicted, it needs to be noted that all of these algorithms produce usable results in the context of working with the glass-forming abilities of metallic glasses.

Despite the slightly better result obtained from the regression tree model, the gradient boosting regression model has an edge over it. Ensemble models better handle the problem of overfitting. This is primarily due to the fact that the ensemble learning paradigm combines the results of various other models to create a more robust one. The resultant model would be more reliable due to its consistency and stability. Nevertheless, it might still be possible to create a better performing model by tuning the selection of hyperparameters. This comparative study aims to uncover the extent to which various ML models suit applications like these.

**Table 1.** List of Machine Learning Algorithms and their respective errors.

| Algorithm | Mean Absolute Percentage Error (MAPE) |
|---|---|
| Linear Regression | 4.64% |
| Bayesian Ridge Regression | 4.61% |
| Gaussian Process Regression | 4.28% |
| Support Vector Regression | 9.40% |
| Decision Tree Regression | 4.04% |
| Gradient Boosting Regression | 4.06% |
| AdaBoost Regression | 4.57% |
| Bagging Regression | 4.37% |

## 6     Conclusion

Based on the results obtained from the proposed architectures, it has been demonstrated that it is possible to effectively predict the reduced glass transition temperature of a material, given the appropriate input features. The scope of this paper does not include the usage of techniques such as deep neural networks and other similar architectures. The primary reason for this restriction is due to the fact that there is no definite procedure for determining the optimal architecture of artificial neural networks. These methods often require an array of specific tuning and pre-processing techniques. However, these methods could be used to further improve the performance of the predictive model in the future.

With more and more advances in the field of computational material science, scientists can reliably base their decisions on conclusions derived from statistics and automated machine learning models. This can fuel the development of better materials which can, in turn, find its place in numerous practical applications.